\documentclass[aps, twocolumn, superscriptaddress]{revtex4-2}
\usepackage{eso-pic,calc}
\usepackage{graphicx}
\usepackage{amsmath, amsthm, amssymb}
\usepackage[caption=false]{subfig}
\usepackage{bm}
\usepackage{ulem}
\usepackage{color}
\usepackage[colorlinks,linktocpage,linkcolor=blue, citecolor=red]{hyperref}
\usepackage{enumitem}
\newtheorem{lemma}{Lemma}[section]

\newtheorem{theorem}[lemma]{Theorem}
\newtheorem{corollary}[lemma]{Corollary}

\theoremstyle{definition}

\def\F{{\cal F}}
\def\Luc{{\cal L}}

\def\prl#1#2#3{{Phys. Rev. Lett.} {\bf #1}, #2 (#3)}
\def\pra#1#2#3{Phys. Rev. A {\bf #1}, #2 (#3)}

\def\pre#1#2#3{Phys. Rev. E {\bf #1}, #2 (#3)}

\def\jpa#1#2#3{J. Phys. A {\bf #1}, #2 (#3)}

\def\jsp#1#2#3{J. Stat. Phys. {\bf #1}, #2 (#3)}

\def\physa#1#2#3{Physica A {\bf #1}, #2 (#3)}

\def\beqr{\begin{eqnarray}}
\def\eqnr{\end{eqnarray}}
\def\beq{\begin{equation}}
\def\bc{\begin{center}}
\def\ec{\end{center}}
\def\eqn{\end{equation}}
\def\bef{\begin{figure}}
\def\enf{\end{figure}}

\def\etl{$et~al.$~}

\begin{document}
\title{The Directed Abelian Sandpile Model on Cylinders}
\author{Abdul Quadir}
\email{abdulq2013@gmail.com}
\affiliation{Mathematics with Computer Science, Guangdong Technion-Israel Institute of Technology, 241 Daxue Road, Shantou, Guangdong Province, 515 063, P. R. China}
\affiliation{Mathematics with Computer Science, Technion-Israel Institute of Technology, Haifa 32000, Israel}
\author{Nikita Kalinin}
\affiliation{Mathematics with Computer Science, Guangdong Technion-Israel Institute of Technology, 241 Daxue Road, Shantou, Guangdong Province, 515 063, P. R. China}
\affiliation{Mathematics with Computer Science, Technion-Israel Institute of Technology, Haifa 32000, Israel}
\author{Ram Ramaswamy}
\affiliation{Department of Physical Sciences, Indian Institute of Science Education and Research, Berhampur, Odisha 760 003, India}

\begin{abstract}
We study the abelian sandpile model in two dimensions on a directed cylindrical lattice with periodic
transverse boundary conditions in the transverse direction and dissipation at one boundary. Recurrent
configurations form a finite abelian group, and repeated grain addition at a specific site generates
deterministic dynamics on this group. Using Dhar's formulation, the sandpile group is identified with
the co-kernel of the reduced directed Laplacian. We show that the group structure admits an exact
reduction to a transverse problem, allowing complete determination of its cyclic decomposition. Our
results establish a direct connection between the algebraic structure of the sandpile group and the
periodicity of the driven dynamics, establishing the manner in which the underlying algebraic structure
governs both deterministic and stochastic evolution in directed sandpile.
\end{abstract}

\maketitle
\section{Introduction}

The abelian sandpile model (ASM)~\cite{BTW, Tang}, a paradigmatic example of self-organized criticality (SOC)
is a slowly driven cellular automaton system that evolves into a critical steady state without fine-tuning of parameters. In the ASM,
local threshold dynamics give rise to scale-invariant avalanches and long-range correlations. A key advance in the
theoretical understanding of the ASM was provided by Dhar \cite{Dhar_1989}, who showed that the set of recurrent
configurations forms a finite abelian group, the so-called sandpile group, whose structure is determined by the reduced
Laplacian of the underlying graph.  These results have been instrumental in making sandpile models and their variants
among the most extensively studied systems in nonequilibrium statistical physics.

The ASM with a preferred direction, namely the directed abelian sandpile model (DASM) is exactly solvable, and its
critical exponents can be computed analytically in all dimensions \cite{Dhar_1989}.
There is a deep connection of the DASM with other nonequilibrium processes such as the voter model~\cite{Dhar},
Scheidegger's river network~\cite{Scheidegger_1967}, and the Takayasu aggregation–diffusion
model~\cite{Takayasu_1989}. Furthermore, the DASM on quasi-one-dimensional geometries such as ladders or strips also
has rich and tractable behavior~\cite{MTZ,Yadav_2012}.

In the present work we consider the DASM in two dimensions where the underlying $n\times L$
lattice has the geometry of a cylinder with $n \gg L$ (hence thin). This generalises the directed ladder
on the $n\times 2$ strip, where it was shown that  the steady-state mass fluctuations can be mapped
exactly to a random walk on a finite ring of size $3^n$~\cite{Yadav_2012}.
The equivalence provides an explicit dynamical interpretation of the sandpile group and connects
SOC behavior with classical stochastic processes. Note that this geometry interpolates between
one-dimensional directed systems and higher-dimensional structures, while retaining sufficient
symmetry to allow for analytical progress, and such results raise the natural question of how
these features extend to higher transverse widths and to geometries with nontrivial topology.
Recent work by Eckmann~\etl~\cite{Eckmann_2023} has also explored sandpile cylinders and
quasi-one-dimensional structures, emphasizing the interplay between geometry and invariant measures.
These studies have primarily focused on the stochastic properties of the dynamics.

\begin{figure}[h]
\centering
\includegraphics[scale=0.1]{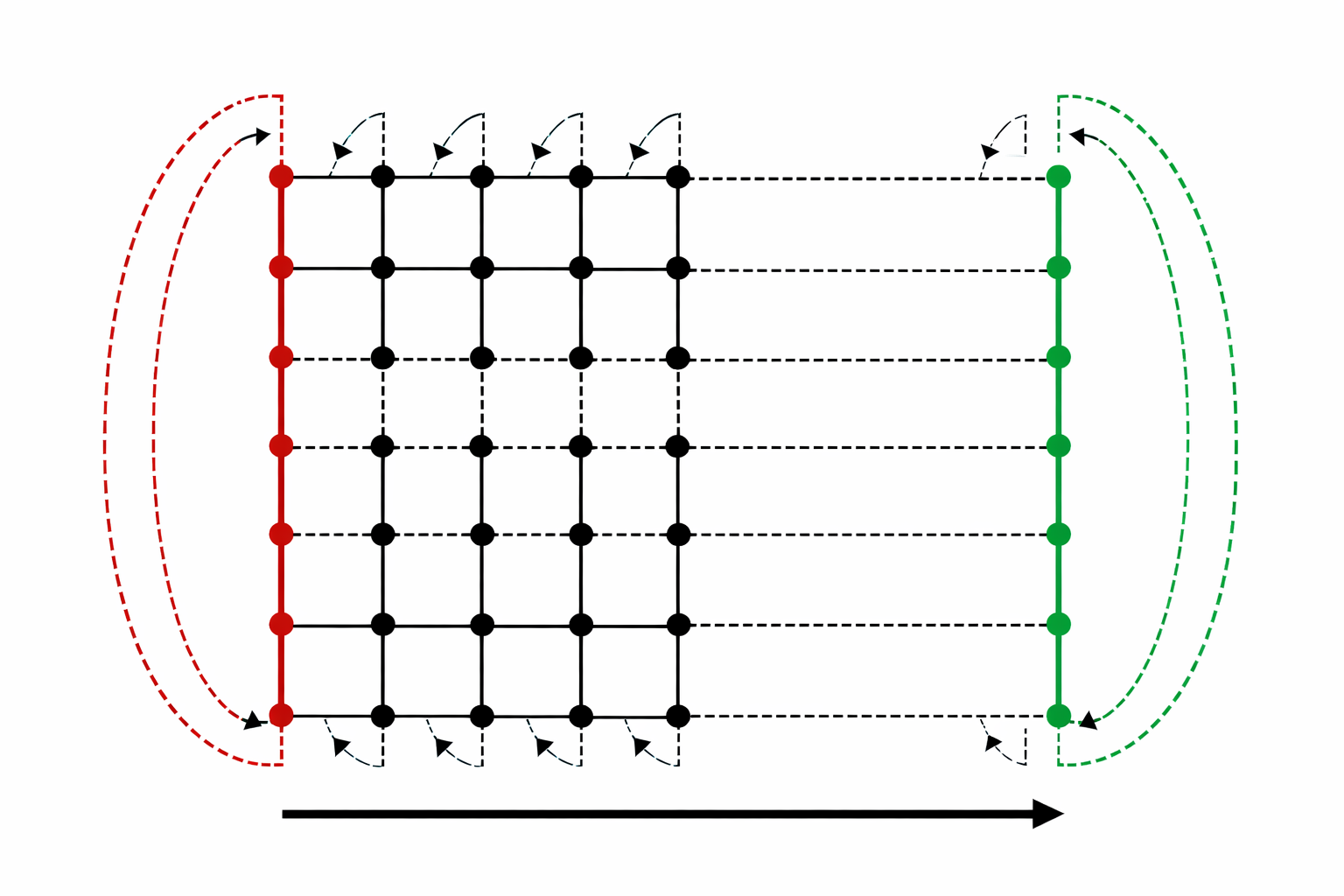}
\caption{Directed cylindrical sandpile of size $n \times L$. Sites are periodic in the transverse ($y$) direction.
Each toppling sends one grain to $(x+1,y)$ and one to $(x,y\pm1)$. Grains leaving at $x=n+1$ are
absorbed at the sink, $s$ (not shown).
The red and green color sites show the driven and dissipation site respectively.}~\label{Fig-Cylinder}
\end{figure}

An important aspect of the ASM, emphasized in Dhar’s formulation, is the algebraic structure
underlying its steady state. The recurrent configurations form a finite abelian group, and the
dynamics of grain addition correspond to deterministic walks on this group. In particular,
repeated addition of grains at a fixed site generates a cyclic evolution whose period is
determined by the order of the corresponding group element. This perspective provides a powerful
link between the microscopic dynamics of the model and the algebraic properties of the associated
Laplacian operator. We express the sandpile group as the
co-kernel of the reduced directed Laplacian. By studying the block structure induced by the
cylindrical geometry, we show that the problem admits an exact reduction to a single transverse block.
This reduction separates the longitudinal and transverse contributions and allows a
complete determination of the sandpile group via the Smith normal form.

As a consequence, we obtain an explicit classification of recurrent configurations
and their cyclic decomposition for cylinders of finite widths. The analysis also provides a direct
connection between the algebraic structure of the sandpile group, the spectral properties
of the transverse operator, and the periodicity of driven dynamics under repeated grain addition.
Our approach provides an exact algebraic characterization of the sandpile group for cylindrical
geometries of arbitrary width $L$. This allows us to systematically determine how the transverse
structure controls both the group decomposition and the resulting dynamical behavior.

The paper is organized as follows. In Sec.~\ref{Sec-ModelDef}, we introduce the reduced Laplacian
for the directed cylindrical sandpile model, and in Sec.~\ref{Sec-SandpileGroup}, show that the sandpile group
is the co-kernel of the Laplacian. We then derive its reduction to a transverse block, and compute the group
structure using the Smith normal form, including an explicit decomposition for finite $L$.
Finally, in Sec.~\ref{Sec-Discussion}  we discuss the dynamical implications of these results, including the
periodicity and the associated random walk, and conclude with a discussion of the broader
significance of the results and possible extensions.

\section{Directed Sandpile on a cylindrical Lattice}~\label{Sec-ModelDef}

Consider a directed sandpile model on a cylindrical lattice of length $n$ and circumference $L$, with longitudinal
coordinate $x=1,\dots,n$, and transverse coordinate $y=1,\dots,L$. The vertex set is
\beq
V=\{(x,y): 1\le x\le n,\ 1\le y\le L\}
\eqn
together with a distinguished sink vertex $s$. The total number of vertices are $|V| = n\cdot L$ +1.
Periodic boundary conditions are imposed in the $y$ direction.
The system is driven by adding grains along the left boundary $x=1$ in two different ways. In the fixed driving case, grains are added repeatedly
at a single boundary site $(1, y_0)$, say at (1,1).
In the random driving case, grains are added at sites $(1,y)$, where $y$ is  chosen randomly in [1,$L$].
For each non-sink vertex $(x,y)\in V$, we introduce three directed edges $(x,y)\to(x,y\pm 1)$ and $(x,y)\to(x+1,y)$,
where $y\pm1$ are taken modulo $L$ [cf. Fig.~\ref{Fig-Cylinder}]. A configuration is a function
\beq
h:V\to\mathbb{Z}_{\ge 0},
\eqn
assigning a non-negative integer height to each site with a vector $\mathbf{h} \in \mathbb{Z}^{nL}_{\geq 0}$. A site $(x,y)$ is
unstable if $h(x,y)\ge 3$ and topples by sending one grain along each outgoing edge
\beqr
&&h(x,y) \mapsto h(x,y)-3,\quad \nonumber \\
&&h(x,y\pm1) \mapsto h(x,y\pm1)+1,~\label{Eq-relax1}
\eqnr
and, for $x<n$,
\beq
h(x+1,y)\mapsto h(x+1,y)+1,~\label{Eq-relax2}
\eqn
while for $x=n$ the forward grain is absorbed by the sink. Thus every non-sink vertex has out-degree 3, and a configuration is stable if
$h(x,y)\le 2$ for all $(x,y)\in V$.
Since each site has a non-negative integer height, the space of all configurations is
\beq
\Omega = \{h: V\to \mathbb{Z}_{\geq 0} \}
\eqn
For stable configuration, each vertices can only have 0,1 or 2 grains.
Since there are $nL$ non-sink sites, the total number of possible configurations is
\beq
|\Omega_{\mathrm{t}}|=3^{nL}.
\eqn
The number of recurrent configurations, namely those that appear in the steady state are a subset, denoted
\beq
\mathcal C(n,L)\subset \Omega_{\mathrm{t}},
\eqn
while the complementary set of forbidden configurations cannot appear in the steady state under repeated addition
and relaxation. The toppling operation can be written in linear form using the reduced Laplacian matrix $\Delta_{n,L}$.
This is an $nL \times nL$ integer matrix indexed by vertices $v,w \in V$, viewed as an $n \times n$ block matrix with
blocks of size $L \times L$. The diagonal entry corresponds to the number of grains lost during toppling, while off-diagonal
entries encode the redistribution to neighboring sites.
With column-wise ordering, $\Delta_{n,L}$ can be written as
\beq
\Delta_{n,L} =
\begin{pmatrix}
B_{L} & -I_{L} & 0 & \cdots & 0 \\
0 & B_{L} & -I_{L} & \ddots & \vdots \\
\vdots & \ddots & \ddots & \ddots & 0 \\
\vdots &  & 0 & B_{L} & -I_{L} \\
0 & \cdots & \cdots & 0 & B_{L}
\end{pmatrix},~\label{Eq-BlockDelta}
\eqn
where $I_{L}$ denotes the $L\times L$ identity matrix, and the diagonal block $B_{L}$ is given by
\beq
B_{L} =
\begin{pmatrix}
3 & -1 & 0 & \cdots & 0 & -1\\
-1 & 3 & -1 & \cdots & 0 & 0\\
0 & -1 & 3 & \ddots & 0 & 0\\
\vdots & \vdots & \ddots & \ddots & -1 & \vdots\\
0 & 0 & 0 & -1 & 3 & -1\\
-1 & 0 & 0 & \cdots & -1 & 3
\end{pmatrix}_{L \times L}.~\label{Eq-BL}
\eqn

The matrix $B_{L}$ describe interactions within each column, while the superdiagonal blocks $-I_{L}$
govern the directed transport from column $x$ to $x+1$. Thus, the block structure reflects the separation
of transverse interactions within columns and directed transport between columns.
It is easy to see that $B_{L}$ is the circulant matrix
\beq
(B_{L})_{i,i}=3,\qquad (B_{L})_{i,i\pm 1}=-1\quad (\text{indices mod }L),
\eqn
and all other entries are $0$. Its determinant $\det(B_L)$ is denoted $D_L$

Following Dhar~\cite{Dhar_1989, Dhar} one can see that recurrent configurations are in one-to-one correspondence with the sandpile group,
\beq
\mathcal C(n,L)\cong \mathcal S(n,L).
\eqn
Hence the number of recurrent configurations is
\beq
|\mathcal C(n,L)|=|\mathcal S(n,L)|=\det(\Delta_{n,L})=D_L^n.
\eqn
While the algebraic structure of the sandpile group is independent of the driving mechanism, the resulting dynamics depends strongly on it. In the
case of fixed driving, repeated addition at a single site generates a deterministic evolution corresponding to iteration of a single group element:
this leads to periodic dynamics in ${\mathcal C(n,L)}$. In contrast, random driving induces a stochastic evolution corresponding to a random walk
on the sandpile group, which explores the space of recurrent configurations.
Thus, deterministic and stochastic dynamics arise from the same underlying algebraic structure.
\section{Sandpile group}~\label{Sec-SandpileGroup}
\subsection{Definition}

A toppling at vertex $v$ corresponds to the update
\beq
\mathbf{h} \longmapsto \mathbf{h} - \Delta^T e_v,
\eqn
where $e_v$ is the standard basis vector in $\mathbb{Z}^{nL}$ corresponding to vertex $v$:
this has entry 1 at the coordinate corresponding to $v$, and $0$ elsewhere. Since each toppling
contributes a vector of the form $-\Delta^T e_v$, any finite sequence of topplings produces a net
change within the integer lattice $\Delta^T \mathbb{Z}^{nL}$. Here $\mathbb{Z}^{nL}$ denotes the
space of configuration vectors, and $\Delta^T \mathbb{Z}^{nL}$ is the sublattice generated by the
columns of $\Delta^T$. Each toppling corresponds to subtracting a column of $\Delta^T$; thus any
sequence of topplings results a linear combination of such columns, and hence lies in the lattice
generated by them.

Two configurations that differ by a sequence of legal topplings (i.e., topplings applied only to unstable sites)
and lead to the same physical state after relaxation are identified modulo $\Delta^T \mathbb{Z}^{nL}$. The
natural quotient space is therefore
\beq
\mathbb{Z}^{nL}/\Delta^T \mathbb{Z}^{nL},
\eqn
which is the co-kernel of $\Delta^T$. Following Dhar's formulation, the sandpile group for the directed
cylinder is defined as
\beq
\mathcal{S}(n,L)=\mathbb{Z}^{nL}/\Delta_{n,L}^{T}\mathbb{Z}^{nL}.
\eqn
Elements of this group correspond to equivalence classes of configurations under toppling, and are in one-to-one
correspondence with recurrent configurations. Since $\Delta_{n,L}$ has full rank, this group is finite and
\beq
|\mathcal{S}(n,L)| = \det(\Delta_{n,L}).
\eqn
While the determinant gives the total number of recurrent configurations, the structure of the Abelian group is obtained via the Smith
normal form of $\Delta_{n, L}^T$. The matrix $\Delta_{n,L}^T$ acts on the lattice $\mathbb{Z}^{nL}$, which denotes the space of
integer-valued configuration vectors indexed by the vertices of the system.

It follows from the Smith normal form theorem for integer matrices~\cite{Dummit_2004book} that there
are uni-modular matrices $U,V \in GL_{nL}(\mathbb{Z})$ such that
\beq
U \Delta_{n,L}^T V = \operatorname{diag}(d_1,\dots,d_{nL}),
\eqn
where $GL_{nL}(\mathbb{Z})$ denotes the group of invertible $nL \times nL$ integer matrices with determinant $\pm 1$,
and the integers $d_i$ satisfy
\beq
d_1 \mid d_2 \mid \cdots \mid d_{nL} \nonumber
\eqn
where the integers $d_i$ are the invariant factors of $\Delta_{nL}^T$. This diagonalization reduces the quotient
\beq
\mathbb{Z}^{nL}/\Delta_{nL}^T \mathbb{Z}^{nL}
\eqn
to a direct sum of cyclic groups, yielding
\beq
\mathcal{S}(n,L) \cong \bigoplus_{i=1}^{nL} \mathbb{Z}/d_i\mathbb{Z}.
\eqn
In practice, factors with $d_i=1$ are trivial and may be omitted.

This reduction makes transparent the role of geometry in the group structure. The longitudinal direction contributes only
through repeated application of the same operator, reflected in the power $n$, while the transverse structure encoded in
the matrix $B_L$ completely determines the invariant factors and hence the cyclic decomposition of the sandpile group.

From a dynamical perspective, each cyclic factor $\mathbb{Z}/d_i\mathbb{Z}$  corresponds to an independent mode of the
evolution, and the order $d_i$ determines a characteristic time scale under repeated grain addition. Thus, the Smith normal
form provides a direct link between the algebraic structure of the sandpile group and the dynamical behavior of the system.

\subsection{Recurrent states in the sandpile group}

Since $B_L$ (see Eq.~\eqref{Eq-BL}) is a circulant matrix, its eigenvalues are 
\beq
\lambda_k = 3 - 2\cos \left( \dfrac{2\pi k}{L} \right),~ k = 0,1,2,\dots ,L-1 ~\label{Eq-Eigenvalues}
\eqn 
and the determinant is
\beq 
D_L = \det(B_L) = \prod_{k=0}^{L-1} \left( 3 - 2\cos\dfrac{2\pi k}{L} \right).~\label{Eq-DetBL}
\eqn 
This can be evaluated [cf. Appendix~\ref{Append-B}], giving
\beq 
D_L = r_+^L + r_-^L - 2,
\eqn 
where $r_\pm = (3 \pm \sqrt{5})/2$. Since $r_+ > 1$ and $0 < r_- < 1$, the determinant grows exponentially with $L$. 
In the large $L$ limit, 
\beq 
D_L \sim r_+^L,
\eqn 
and the total number of recurrent states in the DASM on a cylinder of size $n \times L$ is given by
\beq
|S(n, L)| = \left[ \prod_{k=0}^{L-1} \left( 3 - 2 \cos \dfrac{2 \pi k}{L} \right) \right]^n.
\eqn
For given $L$ this satisfies the first-order recursion
\beq
|\mathcal{S}(n+1, L)| = D_L|\mathcal{S}(n, L)|.
\eqn
Again, from Eq.~\eqref{Eq-DetBL}, the spectrum of $B_L$ is sysmmetric under $k \mapsto L-k$ 
which implies $\lambda_{L-k} = \lambda_k$. The spectral pairing depends on the parity of $L$. 
It can be seen [cf. Appendix~\ref{Append-B}] that the determinant  satisfies the 
recursion
\beq 
D_L = 3D_{L-1} - D_{L-2} + 2, ~~~ L \geq 3
\eqn 
with initial conditions $D_1=1,~D_2=5$. 

\subsection{Driving at a fixed site: Periodic orbits in configuration space}
Let $\mathcal{C}$ denote the set of recurrent configurations, and let $R \in \mathcal{C}$ be a
recurrent configuration. We define the addition operator $A_{1,y}$ as the map that adds a grain at
site $(1,y)$ and stabilizes the resulting configuration. Start from a recurrent state $\mathcal{C}$,
applying $A(1, y)$ and relaxing to a recurrent state again is the deterministic map
\beq
A_{1,y=1}: R \mapsto R;~~ A_{1,y=1} (\mathcal{C}) \mapsto \it{Stabilize}(\mathcal{C} + e_{1,y=1})
\eqn
where {\it Stabilize} denotes the relaxation process that maps a configuration to its unique stable (and recurrent) representative via topplings and the notation $[e_{1,y}]$ denotes the equivalence class of the configuration corresponding to adding one grain at $(1,y)$ in the sandpile group $\mathcal{S}(n,L)$. On recurrent states, $A_{1,y=1}$ is just translation by one group element,
\beq
A_{1,y=1} (\mathcal{C}) = \mathcal{C} \oplus [e_{1,y=1}]
\eqn
so after $m$ steps,
\beq
A^m_{1,y=1} (\mathcal{C}) = \mathcal{C} \oplus m[e_{1,y=1}]
\eqn
Here $\oplus$ denotes addition followed by stabilization, which defines the Abelian group operation on recurrent configurations. Therefore,
\beq
A^m_{1,y=1}(\mathcal{C}) = \mathcal{C} \iff m[e_{1,y=1}] = 0 \text{~in~} \mathcal{S}(n,L)
\eqn
So the period does not depend on which recurrent state $\mathcal{C}$ starts from. It depends only on the group element $[e_{1,y=1}]$. So the period is
\beq
per(1,1) = ord([e_{1,1}])
\eqn
where $ord(g)$ denotes the order of the group element $g$, i.e., the smallest positive integer m such that $mg=0$ in $\mathcal{S}(n,L)$.

Thus a fixed driving site generates a cyclic subgroup of the sandpile group, and the dynamics is confined to a single orbit determined by the driving site. For the ladder case $L$ = 2, this reduces to a single cyclic group of size $3^n$, recovering the exact periodic
behavior reported in~\cite{Yadav_2012}. For higher widths, the presence of multiple invariant factors leads to
a richer set of possible periods [cf. Appendix~\ref{Append-B}.2].

\subsection{Driving at random sites: Random walk on the sandpile group}
If the driving is at a random site on the boundary, then in general there is no single deterministic periodic orbit
and the evolution is no longer a deterministic map. Instead we have a random walk on the sandpile
group generated by the elements $g_y=[e_{1,y}]$, where $y \in \{1,2,\dots,L \}$ is chosen randomly.
Adding a grain at the randomly chosen site and subsequently stabilizing the configuration corresponds to addition of the group element $g_y$ in the sandpile group.
This defines a Markov chain on $\mathcal{S}(n,L)$ with transition probabilities determined by the choice of $y$ with equal probability.
Let $Y_t$ denote the random variable specifying the chosen driving site at time $t$. The evolution is therefore given by
\beq
\mathcal{C}_{t+1}=\mathcal{C}_t \oplus g_{Y_t},
\eqn
where $g_{Y_t}$ denotes the corresponding randomly chosen generator.
If the generators $\{ g_y \}^L_{y=1}$ generate the full sandpile group, the Markov chain is irreducible and the stationary distribution is uniform over recurrent configurations. For a finite irreducible Markov chain with uniform stationary distribution, the mean return time to a recurrent configuration equals the total number of recurrent configurations. This follows from Kac's recurrence theorem for finite Markov chains~\cite{Kac_1947}.

For fixed cylinder length $n$, the size of the recurrent state grows exponentially 
with the width $L$, namely
\beq
|\mathcal{S}(n,L)| = D_L^n \sim r_+^{\,nL}, \qquad r_+ = (3+\sqrt5)/2.
\eqn
For the case $n=1$, namely a ``directed'' ring, we have 
\beq
|\mathcal{S}(1,L)| \sim e^{b_1L},
~~ b_1=\log r_+.
\eqn

\begin{figure}[h]
    \centering
    \includegraphics[scale=0.75]{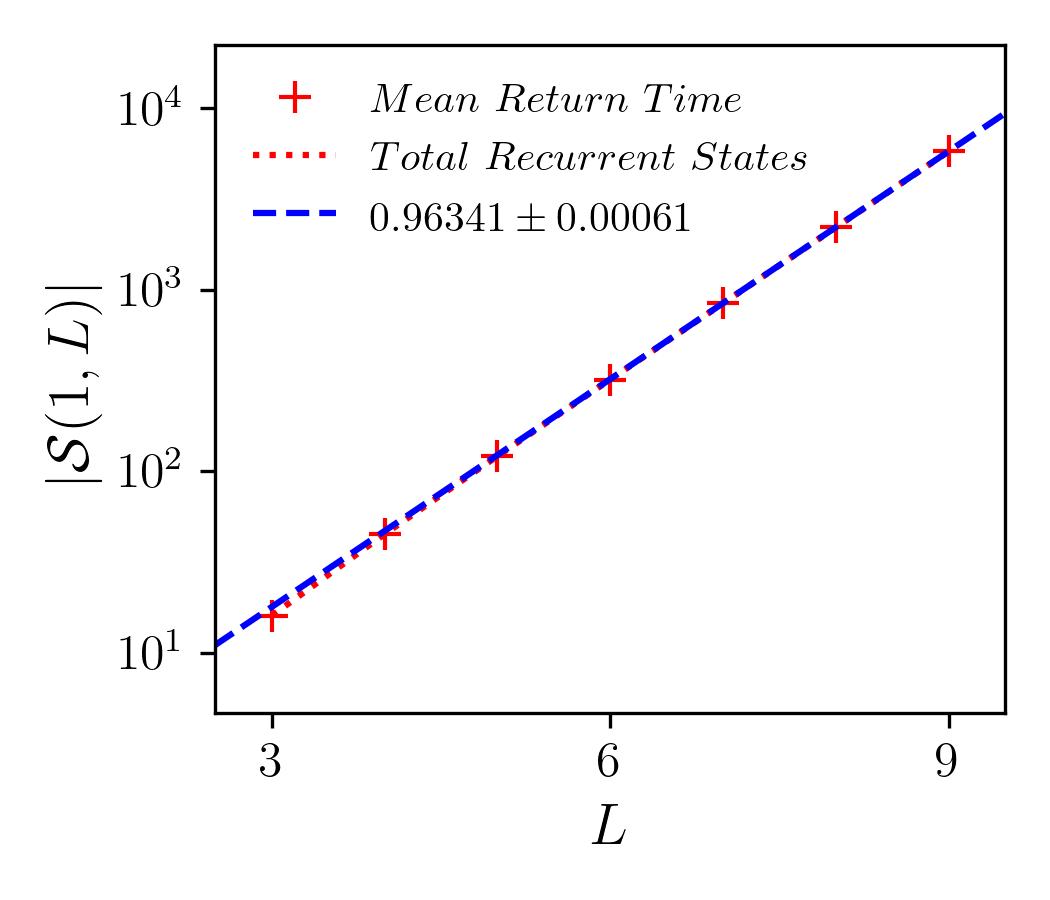}
    \caption{Exponential growth of the mean return time (equivalently, the number of recurrent configurations) with the transverse width $L$ for $n=1$. The dashed line shows the exponential fit $|\mathcal{S}(1,L)|\sim e^{b_1L}$, where the fitted exponent is close to the theoretical value $b_1=\log(r_+)=0.96242$.}~\label{Fig-Total-Return-Time}
\end{figure}

The exponential behaviour is verified numerically by estimating the average mean return time for random driving within the recurrent configuration subspace. 
For each value of $L$, the dynamics is evolved for $10^8$ driving steps, and the return statistics are averaged over the generated trajectory. 
Since the stationary distribution is uniform, the measured mean return time provides a direct numerical estimate of $|\mathcal{S}(1,L)|$. 
Our results are shown in Fig.~\ref{Fig-Total-Return-Time} where this quantity is plotted as a function of $L$, and the extracted exponent 
is $b_1 = 0.96341 \pm 0.00061$, which is in good agreement with the exact theoretical prediction $\ln(r_+)$ = 0.96242\ldots. 
We have also carried out simulations for $n=2, 3$ and find exponential growth in those cases as well.

The Smith normal form also provides a direct way to extract the characteristic time scales of the dynamics. By writing the sandpile group
as a direct sum of cyclic components, each invariant factor $d_i$ corresponds to an independent dynamical mode. Under fixed driving at a
site $(1,y)$, the evolution corresponds to repeated addition of a group element $g=[e_{1,y}]$, whose components in the Smith normal
form basis determine the periodic behavior. The system returns to its initial state when $mg=0$, which requires $mg_i \equiv 0~(\mod~d_i)$
for all $i$. Here, $g_i$ denotes the component of $g$ in the Smith normal form basis.
This yields characteristic time scales $m_i=\dfrac{d_i}{gcd(d_i, g_i)}$, and the overall period is given by the least common multiple of these.

Such random walks on the sandpile group provide a natural framework to understand temporal correlations and spectral properties of
the dynamics. In particular, previous studies have shown that these dynamics can give rise to $1/f^\alpha$ noise in suitable
observables~\cite{MTZ, Yadav_2012}, linking the algebraic structure to experimentally relevant fluctuation behavior.

In the case of random driving, all generators contribute, and the dynamics involves multiple time scales associated with the invariant
factors. This multi-scale structure can contribute to rich temporal behavior and is closely related to the emergence of long-range correlations
and $1/f^\alpha$ noise observed in sandpile dynamics~\cite{Yadav_2012, MTZ, Eckmann_2023}. Thus, both deterministic periodicity and stochastic fluctuations emerge from the same underlying algebraic structure of the sandpile group.

\section{Discussion}~\label{Sec-Discussion}
We have analyzed the directed abelian sandpile model on a cylindrical geometry and obtained an exact characterization of the sandpile group.
By studying the block structure of the Laplacian, we reduced the problem to a transverse matrix, allowing an explicit determination of the
group via its Smith normal form. This extends the algebraic framework introduced in the seminal work by Dhar and Ramaswamy~\cite{Dhar_1989},
where the sandpile group was first identified as the co-kernel of the Laplacian, by providing an explicit classification for a nontrivial
cylindrical geometry.

The present results provide a direct bridge between the algebraic structure of the sandpile group and the dynamical properties of directed
sandpile models on cylindrical geometries. In particular, the decomposition of the sandpile group determines the possible periodicity
under repeated grain addition, providing a direct link between algebraic invariants and dynamical behavior. Thus, both deterministic
periodicity and stochastic fluctuations arise from the same underlying group structure.
For fixed driving, the dynamics is confined to cyclic subgroups generated by the corresponding addition operators, while random driving produces an ergodic random walk on the sandpile group whenever the generators span the full group. The invariant factors obtained from the Smith normal form therefore introduce characteristic algebraic time scales that can influence temporal correlations and spectral properties.
This complements earlier studies of dynamical fluctuations and spectral properties in directed sandpile~\cite{Dhar_2016}, where temporal
correlations and $1/f^\alpha-$noise were analyzed. The ladder geometry appears as a special case within this framework, recovering known
results and illustrating the generality of the approach.
The ladder geometry appears as a special case within this framework, recovering the cyclic structure associated with the random-walk formulation studied previously~\cite{Yadav_2012}. Our results show that the ladder model arises naturally as the $L=2$ limit of the cylindrical construction.
Our formulation shows that this behavior arises naturally as a limiting case of the cylindrical construction.

The present work suggests that directed sandpile on quasi-one-dimensional geometries admit a systematic classification in terms of their
transverse block operators. Related approaches, such as the analysis of invariant measures and dynamical structures on cylindrical graphs~\cite{MTZ},
further highlight the role of geometry in determining the statistical properties of sandpile models.

It would be interesting to extend these results to more general directed graphs and to further explore the interplay between sandpile groups, spectral structures, stochastic dynamics, and non-equilibrium fluctuations.



\section{Acknowledgment}
A. Q. is grateful to Jean-Pierre Eckmann and Tatiana Smirnova-Nagnibeda for their kind invitation to the University of Geneva
and for the insightful discussions during his stay. He further thanks Deepak Dhar for valuable conversations at ISPCM 2025, ICTS Bengaluru.

\appendix

\section{Closed-form evaluation of $D_L$}~\label{Append-B}

\begin{theorem}
    Since $B_L = 3I - A_{C_L}$, where $A_{C_L}$ is the adjacency matrix of the cycle graph $C_L$. Then
    \beq
    D_L = \left(\frac{3+\sqrt5}{2}\right)^{L} + \left(\frac{3-\sqrt5}{2}\right)^{L} -2.~\label{Thoerem1-B}
    \eqn
\end{theorem}
\noindent
Using the eigenvalue representation, we define
\beq
P_{L} := \prod_{k=0}^{L-1} \left( 3- 2 cos \dfrac{2\pi k}{L} \right)
\eqn
Since $\det{B_L}$ equals the product of eigenvalues, we have $D_L= P_L$.

\noindent
Let $z_k = e^{2\pi i k / L},~ k=0,1,\dots,L-1$, then
\beq
3 - 2\cos\left(\frac{2\pi k}{L}\right) = 3 - z_k - z_k^{-1}
\eqn
Now define
\beq
r_\pm = \frac{3 \pm \sqrt{5}}{2}
\eqn
which are the roots of the equation $r^2 - 3r +1 = 0$. Hence
\beq
(z_k^2 - 3z_k + 1) = (z_k - r_+) (z_k - r_-) \nonumber
\eqn
and therefore,
\beq
3 - z_k - z_k^{-1} = -\dfrac{(z_k - r_+)(z_k - r_-)}{z_k}. \nonumber
\eqn
Substituting into $P_L$
\beqr
P_{L} &=& \prod_{k=0}^{L-1} \left(3 - z_k - z_k^{-1}\right) \nonumber \\
    &=& (-1)^L \frac{ \prod_{k=0}^{L-1}(z_k - r_+) \prod_{k=0}^{L-1}(z_k - r_-)}{\prod_{k=0}^{L-1} z_k}~\label{Eq-Form-PL}
\eqnr
Since $z_k$ are the roots of $z^{L} - 1 = 0$, we have the identity
\beq
\prod_{k=0}^{L-1} (r - z_k) = r^L - 1.
\eqn
Thus,
\beq
\prod_{k=0}^{L-1} (z_k - r) = (-1)^L (r^L - 1).
\eqn
Applying this identity for $r=r_{\pm}$,
\beq
\prod_{k=0}^{L-1} (z_k - r_\pm) = (-1)^L (r_\pm^L - 1), \nonumber
\eqn
Furthermore,
\beq
\prod_{k=0}^{L-1} z_k = e^{2\pi i \frac{0+1+\cdots+(L-1)}{L}} = e^{\pi i (L-1)} = (-1)^{L-1}. \nonumber
\eqn
Substitute these expressions into the Eq.~\eqref{Eq-Form-PL}
\beqr
P_L = -(r_+^{L} r_-^{L} - r_+^{L} - r_-^{L} + 1) \nonumber
\eqnr
Since $r_+r_-=1$ follows $r_+^{L} r_-^{L} = 1$, hence
\beq
P_{L} = r_+^{L} + r_-^{L} - 2.
\eqn
Therefore,
\beq
D_L = \left(\frac{3+\sqrt{5}}{2}\right)^L + \left(\frac{3-\sqrt{5}}{2}\right)^L -2.
\eqn

\paragraph{Case $L=3$}
The eigenvalues are $1,4,4$, and
\beqr
D_3 &=& \left( \dfrac{3+\sqrt{5}}{2} \right)^3 + \left( \dfrac{3-\sqrt{5}}{2} \right)^3 - 2 \nonumber\\
&=& 16= 1\cdot 4\cdot 4.
\eqnr

\paragraph{Case $L=4$}
Since the eigenvalues are 1,3,5,3,
\beqr
D_4
&=&\left( \dfrac{3+\sqrt{5}}{2} \right)^4 + \left( \dfrac{3-\sqrt{5}}{2} \right)^4 - 2\nonumber\\
&=& 45 = 1\cdot 3\cdot 5\cdot 3.
\eqnr

\begin{corollary}{Recurrence relation for $D_L$}
    The determinant $D_L$ satisfies the recurrence relation
    \beq
    D_L = 3D_{L-1} - D_{L-2} + 2, \qquad L \ge 3,
    \eqn
    with initial conditions $D_1=1,~D_2=5$.
\end{corollary}

Let $r_\pm = (3 \pm \sqrt{5})/2$ be the roots of the quadratic equation
\beq
r^2 - 3r + 1 = 0.
\eqn
Define the sequence $a_L := r_+^{L} + r_-^{L}$. Since $r_\pm$ satisfy the characteristic equation $r^2-3r+1=0$ and the sequence $a_L$ obeys the linear recurrence
\beq
a_L = 3a_{L-1} - a_{L-2}.
\eqn
From Theorem~\ref{Thoerem1-B}, the determinant can be written as
\beq
\det(B_L) = a_L - 2,
\eqn
Hence,
\beq
D_L = 3D_{L-1} - D_{L-2} + 2, \qquad L \ge 3,
\eqn
with initial values $D_1 = 1,~ D_2 = 5$. 


\section{Algebraic structure of the sandpile groups 
on a directed cylinder}
Using the reduction
\beq 
\mathcal{S} = \mathbb{Z}^L/B^n_L \mathbb{Z}^L,
\eqn 
the structure of the sandpile group is completely determined by the Smith normal form of the transverse matrix $B_L^n$. 
Thus, the transverse geometry controls the cyclic decomposition of the sandpile group, while the longitudinal direction 
contributes through the power $n$. The arithmetic properties of $B_L$ lead to a parity-dependent structure closely 
related to Fibonacci and Lucas numbers $\F_L$ and $\Luc_L$. The Fibonacci numbers satisfy the recursion $\F_{L}=\F_{L-1}+\F_{L-2}$ 
with initial conditions $\F_0=0,\F_1=1$, while the Lucas numbers have a similar recursion, 
$\Luc_{L}=\Luc_{L-1}+\Luc_{L-2}$ but with initial conditions $\Luc_0=2, \Luc_1=1$.\\

\subsection{Parity structure}
There is  strong parity dependence of both $D_L$ and the invariant-factors of the transverse block $B_L$. 
In particular, odd and even transverse widths lead to qualitatively different algebraic structures. 
This distinction originates from the arithmetic structure of the eigenvalues of $B_L$ and their factorization. 

\noindent
\textbf{(i) Odd $L$.} For odd transverse width $L$, the determinant satisfies
\beq 
D_L = \Luc_L^2,
\eqn 
and the sandpile group takes the simple form
\beq 
\mathcal{S}(n,L) \cong (\mathbb{Z}/\Luc_L^n \mathbb{Z})^2.
\eqn 
Therefore, 
\beq 
|\mathcal{S}(n,L)| = \Luc_L^{2n},
\eqn 
and  for odd transverse widths, the sandpile group decomposes into two identical cyclic components. For small $L$ odd, we get
\beqr
\mathcal{S}(n,3) &\cong & (\mathbb{Z}/4^n \mathbb{Z})^2, \quad \mathcal{S}(n,5) \cong (\mathbb{Z}/11^n \mathbb{Z})^2, \nonumber \\
\mathcal{S}(n,7) &\cong & (\mathbb{Z}/29^n \mathbb{Z})^2, \quad \mathcal{S}(n,9) \cong  (\mathbb{Z}/76^n \mathbb{Z})^2. \nonumber
\eqnr 

\noindent
\textbf{(ii) Even $L$.} For even transverse width $L$, the structure is governed by the Fibonacci numbers. In this case,
\beq 
D_L = 5\F_L^2
\eqn
\noindent
For $n=1$, namely the ring, the decomposition is
\beq 
\mathcal{S}(1, L) \cong \mathbb{Z}/\F_L \mathbb{Z} \oplus \mathbb{Z}/(5\F_L) \mathbb{Z},
\eqn
while for larger $n$, the structure becomes more intricate. If
\beq 
\gcd(\F_L, L) =  1,
\eqn
the decomposition simplifies to
\beq 
\mathcal{S}(n, L) \cong \mathbb{Z}/F^n_L \mathbb{Z} \oplus \mathbb{Z}/(5^nF^n_L) \mathbb{Z}.
\eqn
Otherwise, additional cyclic factors may appear, and the full decomposition must be obtained from the Smith normal form of $B^n_L$. 
For small even widths $L$, we get
\beqr 
\mathcal{S}(n, 4) &\cong & \mathbb{Z}/3^n \mathbb{Z} \times \mathbb{Z} / 15^n \mathbb{Z}, \nonumber \\
\mathcal{S}(n, 6) &\cong & (\mathbb{Z}/2^{n-1} \mathbb{Z})^2 \times \mathbb{Z}/2^{2n+1} \mathbb{Z} \times \mathbb{Z}/(2^{2n+1}5^n)\mathbb{Z}, \nonumber \\
\mathcal{S}(n, 8) &\cong & \mathbb{Z}/21^n \mathbb{Z} \times \mathbb{Z}/105^n \mathbb{Z}. \nonumber
\eqnr 
In contrast to the odd-width case, even transverse widths generally produce asymmetric invariant factors and may contain additional cyclic components.

\subsection{The ladder $L=2$ without periodicity in $y$.}
The directed ladder \cite{MTZ,Yadav_2012}  is a special limit of the cylindrical construction. The threshold is set to 2 rather than 3, 
and toppling is to one transverse neighbour and one particle is transferred longitudinally. The corresponding transverse block is
\beq 
B_{\mathrm{lad}}=\begin{pmatrix}2&-1\\-1&2\end{pmatrix},
\qquad \det(B_{\mathrm{lad}})=3,
\eqn
and the same block structure holds with $B_{\mathrm{lad}}$ in place of $B_{L}$. Then
\beq
\mathcal{S}_{\mathrm{lad}}(n)\cong \mathbb{Z}^2/B_{\mathrm{lad}}^{\,n}\mathbb{Z}^2\cong \mathbb{Z}/3^n\mathbb{Z}.
\eqn
This shows that the ladder $L=2$ fits naturally within the present formulation. The $2\times n$ ladder appears as a 
limiting case of the cylindrical construction, with a single nontrivial invariant factor~\cite{Yadav_2012}.
The reduction thus yields a cyclic group of order $3^n$.


\end{document}